\documentclass{sigchi}

\toappear{\scriptsize Permission to make digital or hard copies of all or part of this work for personal or classroom use is granted without fee provided that copies are not made or distributed for profit or commercial advantage and that copies bear this notice and the full citation on the first page. Copyrights for components of this work owned by others than ACM must be honored. Abstracting with credit is permitted. To copy otherwise, or republish, to post on servers or to redistribute to lists, requires prior specific permission and/or a fee. Request permissions from permissions@acm.org. \\
{\emph{CSCW'14}}, February 15--19, 2014, Baltimore, Maryland, USA. \\
Copyright \copyright~2014 ACM 978-1-4503-2540-0/14/02...\$15.00. \\
http://dx.doi.org/10.1145/2531602.2531642 }
\usepackage{balance}  % to better equalize the last page
\usepackage{graphics} % for EPS, load graphicx instead
\usepackage{times}    % comment if you want LaTeX's default font
\usepackage{url}      % llt: nicely formatted URLs
\usepackage{cite}

\makeatletter
\def\url@leostyle{%
  \@ifundefined{selectfont}{\def\UrlFont{\sf}}{\def\UrlFont{\small\bf\ttfamily}}}
\makeatother
\def\pprw{8.5in}
\def\pprh{11in}

\newcommand{\omt}[1]{}
\newcommand{\xhdr}[1]{\paragraph*{\bf #1.}}

\def\rs{\vspace*{-0.10in}}

\begin{document}

\title{Romantic Partnerships and the Dispersion of Social Ties: \\ A Network Analysis of Relationship Status on Facebook
% \thanks{
% This research has been
% supported in part by grants from
% the Simons Foundation and from
% the National Science Foundation.
% }
}

% \numberofauthors{1}
% \author{
  % \alignauthor Authors Anonymized for Conference Submission \\
 % }

 \numberofauthors{2}
 \author{
  \alignauthor Lars Backstrom\\
    \affaddr{Facebook Inc.}\\
     % \email{lars@fb.com}\\
   \alignauthor Jon Kleinberg\\
     \affaddr{Cornell University}\\
     % \email{kleinber@cs.cornell.edu}\\
 }

% Teaser figure can go here
%\teaser{
%  \centering
%  \includegraphics{Figure1}
%  \caption{Teaser Image}
%  \label{fig:teaser}
%}

\maketitle

\begin{abstract}

A crucial task in the analysis of on-line social-networking systems
is to identify important people --- those linked by strong social ties ---
within an individual's network neighborhood.
Here we investigate this question for a particular category of
strong ties, those involving spouses or romantic partners.
We organize our analysis around a basic question:
given all the connections among a person's friends,
can you recognize his or her romantic partner from the network structure alone?
Using data from a large sample of Facebook users,
we find that this task can be accomplished with high accuracy,
but doing so requires the development of a new measure of tie strength
that we term `dispersion' --- the extent to which two people's mutual
friends are not themselves well-connected.
The results 
offer methods for identifying types
of structurally significant people in on-line applications, and
suggest a potential expansion of existing theories
of tie strength.

\end{abstract}

% \omt{
\vspace{0.05in}
\noindent
{\bf Categories and Subject Descriptors:}
H.2.8 [{\bf Database Management}]: Database applications---{\em Data mining}
%\category{G.2.2}{Mathematics of Computing}{Graph Theory}
%A category including the fourth, optional field follows...
%\category{D.2.8}{Software Engineering}{Metrics}[complexity measures, performance measures]
%\terms{Theory}

\vspace{0.0in}
\noindent
{\bf Keywords:}
%\keywords{
Social Networks; Romantic Relationships.
%}
\vspace{0.0in}
% }

% \textcolor{red}{The following section is mandatory for accepted papers, but not needed for submissions for the June 1 deadline.}
% \keywords{
	% Guides; instructions; author's kit; conference publications;
	% keywords should be separated by a semi-colon.
% }
% 
% \category{H.5.m.}{Information Interfaces and Presentation (e.g. HCI)}{Miscellaneous}
% 
% See: \url{http://www.acm.org/about/class/1998/}
% for more information and the full list of ACM classifiers
% and descriptors. 
% 
% \textcolor{red}{The following section can be included for accepted papers, but is not needed for submissions for the June 1 deadline.}
% 
% \terms{
	% Theory; Measurement. 
	% If you choose more than one ACM General Term, 
	% separate the terms with a semi-colon.
% }

% See list of the limited ACM 16 terms in the
% instructions and additional information:
% \url{http://www.sheridanprinting.com/sigchi/generalterms.htm}.
% \textcolor{red}{Optional section to be included in your final version.}
% 

\section{Introduction}

In a social network, 
an individual's network neighborhood ---
the set of people to whom he or she is linked ---
has been shown to have important consequences in a wide range of
settings, including social support
\cite{fischer-dwell-book,mcpherson-impt-matters}
% \cite{fischer-dwell-book,mcpherson-impt-matters,wellman-wortley-support}
and professional opportunities
\cite{granovetter-weak-ties,burt-struct-holes-book}.
% \cite{granovetter-weak-ties,burt-struct-holes-book,burke-losing-job-fb}.
As people use on-line social networks to manage increasingly
rich aspects of their lives, the structures of their on-line
network neighborhoods have come to reflect these functions, and
the complexity that goes with them.
% \blfootnote{
% \noindent This research has been
% supported in part by grants from
% the Simons Foundation and from
% the National Science Foundation.
% }

A person's network neighbors, taken as a whole, 
encompass a profoundly diverse set of relationships ---
they typically include family members, co-workers, friends of long duration,
distant acquaintances, potentially a spouse or romantic partner, and
a variety of other categories.
An important and very broad issue for the analysis of on-line social networks
is to use features in the available data to recognize this variation across 
types of relationships.
Methods to do this effectively can play an important role
for many applications at the interface between an individual
and the rest of the network ---
managing their on-line interactions
\cite{farnham-social-facets},
prioritizing content they see from friends \cite{backstrom-wsdm13},
and organizing their neighborhood into conceptually coherent groups
\cite{mcauley-social-circles,min-social-roles-cscw13}.
% \cite{farnham-social-facets,mcauley-social-circles,min-social-roles-cscw13}.

\xhdr{Tie Strength}
{\em Tie strength} forms an important dimension along which 
to characterize a person's links to their network neighbors.
Tie strength informally refers to the `closeness' of a friendship;
it captures a spectrum that ranges from
strong ties with close friends to
weak ties with more distant acquaintances.
An active line of research reaching back to foundational work
in sociology has studied the relationship between
the strengths of ties and their structural role in the underlying
social network \cite{granovetter-weak-ties}.
Strong ties are typically `embedded' in the network,
surrounded by a large number of mutual friends
\cite{granovetter-embeddedness,coleman-social-capital},
% \cite{granovetter-embeddedness,coleman-social-capital,marsden-tie-strength},
% burt-struct-holes-book,
% gilbert-tie-strength,
% kossinets-email,
% shi-strong-ties,
% onnela-phone-data},
and often involving large amounts of shared time together
\cite{marsden-tie-strength}
and extensive interaction
\cite{jones-tie-strength}.
% \cite{marsden-tie-strength,eagle-inferring-friendship,jones-tie-strength}.
Weak ties, in contrast, often involve few mutual friends and
can serve as `bridges' to diverse parts of the network,
providing access to novel information
\cite{granovetter-weak-ties,burt-struct-holes-book}.
% \cite{granovetter-weak-ties,granovetter-swt-revisited,burt-struct-holes-book}.

A fundamental question connected to our understanding of strong ties
is to identify the most important individuals
in a person's social network neighborhood using the underlying 
network structure.  
What are the defining structural signatures of a person's strongest ties,
and how do we recognize them?
Techniques for this problem have potential importance
both for organizing a person's network neighborhood in on-line applications,
and also for providing basic insights into the effect of close
relationships on network structure more broadly.

Recent work has developed methods of analyzing and
estimating tie strength in on-line domains, drawing on data from
e-mail \cite{kossinets-email},
phone calls \cite{onnela-phone-data},
and social media \cite{gilbert-tie-strength}.
% \cite{gilbert-tie-strength,kossinets-email,shi-strong-ties,onnela-phone-data}.
The key structural feature used in these analyses is the notion
of {\em embeddedness} --- the number of mutual friends two people share
\cite{marsden-tie-strength},
% \cite{granovetter-embeddedness,marsden-tie-strength},
a quantity that typically increases with tie strength.
Indeed, embeddedness has been so tightly associated with tie strength that
it has remained largely an open question to determine whether there
are other structural measures, distinct from embeddedness,
that may be more appropriate for characterizing particular types of strong ties.

\xhdr{Romantic Relationships}
In this work we propose a new network-based characterization
for intimate relationships, those involving spouses or
romantic partners. 
Such relationships are important to study for several reasons.

From a substantive point of view, romantic relationships
are of course singular types of
social ties that play powerful roles in social processes 
over a person's whole life course
\cite{bott-family-soc-net},
from adolescence \cite{bearman-dating} to
older age \cite{cornwell-spousal-overlap}.
% \cite{bearman-dating,bott-family-soc-net,cornwell-spousal-overlap}.
They also form an important aspect of the everyday practices and 
uses of social media
\cite{zhao-rel-partner-fb}.
% \cite{bowe-rel-partner-fb,donath-public-displays,muise-fb-jealousy,utz-rel-partner-fb,zhao-rel-partner-fb}.
And they are an important challenge from a methodological point of view;
they are evidently among the very strongest ties,
but it has not been clear whether standard structural theories
based on embeddedness are sufficient to characterize them, or
whether they possess singular structural properties of their own
\cite{felmlee-rel-partner,kalmijn-shared-networks}.

Our central finding is that embeddedness is in fact
a comparatively weak means of characterizing romantic relationships, 
and that an alternate network measure that we term 
{\em dispersion} is significantly more effective.
Our measure of dispersion looks not just at the number of
mutual friends of two people, but also at the network structure
on these mutual friends; roughly, a link between two people 
has high dispersion when their mutual friends are not well connected
to one another.

On a large random sample of Facebook users who have declared 
a relationship partner in their profile, we find that our 
dispersion measure has roughly twice the accuracy of embeddedness
in identifying this partner from among the user's full
set of friends.
Indeed, for married Facebook users, our measure of dispersion 
applied to the pure, unannotated network structure is more
effective at identifying a user's spouse than a complex classifier
trained using machine learning on an array of interaction measures
including messaging, commenting, profile-viewing, and co-presence 
at events and in photos.
Further, using dispersion in conjunction with these interaction features
produces significantly higher accuracy.
% produces a classifier with significantly higher accuracy.

The main contributions of our work are thus the following.
\begin{itemize}
\item We propose a new network measure, {\em dispersion},
for estimating tie strength.  Given the ubiquity of embeddedness
in existing analyses of tie strength, the availability of this new measure
broadens the range of tools available for reasoning about tie strength,
and about mechanisms for tie-strength classification in on-line domains.
% It may thus be useful for a range of personal network management tasks
% that involve characterizing different types of relationships,
% as noted above.
\item We provide a new substantive characterization of 
romantic relationships in terms of network structure, 
with potential consequences for our
understanding of the effect that such relationships have
on the underlying social network.
\item Given this characterization, we examine its variation
across different conditions and populations.  We find, for example,
that there are significant gender differences in the extent to which
relationship partners are recognizable from network structure, 
and that relationships are more likely to persist when they score highly
under our dispersion measure.
\end{itemize}

It is also important to delineate the scope of our results.
Our approach to analyzing romantic partnerships in
on-line social settings is through their effect
on the network structure, and the ways in which 
such relationships can be recognized through their structural 
signatures.  As such, there is potential for it to be
combined with other perspectives on how
these relationships are expressed on-line, and the conventions that
develop around their on-line expression; a complete picture will
necessarily involve a synthesis of all these perspectives.

\section{Data and Problem Description}

\begin{figure}[t]
 \begin{center}
 \includegraphics[width=0.37\textwidth]{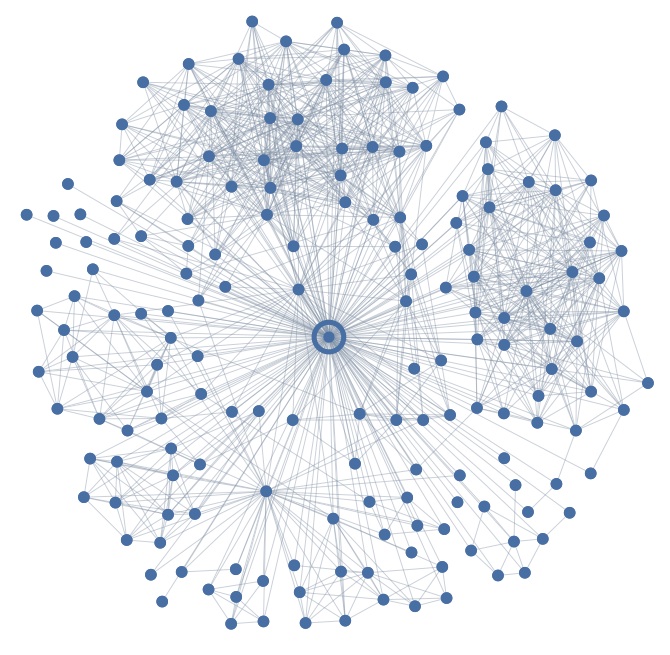}
 \end{center}
\caption{
 \label{fig:nb-real}
% \footnotesize
A network neighborhood, contributed by 
a Facebook employee (drawn as the circled node at the center),
and displayed as an example in the work of 
Marlow et al\protect\cite{marlow-facebook-tie-strength-ext}.
Two clear clusters with highly embedded links are visible at the top
and right of the diagram; in the lower left of the diagram are
smaller, sparser clusters together with a node
that bridges between these clusters.
\rs 
  }
\end{figure}

We analyze romantic relationships in social networks
using a dataset of randomly sampled Facebook users
who declared a relationship partner in their profile;
this includes users who listed their status as 
`married,' `engaged,' or `in a relationship'.
% \cite{note-rel-partner01}.
To evaluate different structural theories on a common footing, we begin with
a simply stated prediction task designed to capture the basic issues.
We take a Facebook user with a declared relationship partner,
and we hide the identity of this partner.
Then we ask: given the user's network
neighborhood --- the set of all friends and the links among them ---
how accurately can we identify the relationship partner
using this structural information alone?
Figure \ref{fig:nb-real} gives an example of such a Facebook user's
network neighborhood\cite{marlow-facebook-tie-strength-ext},
drawn so that the user is depicted at the center;
such diagrams are the `input' from which we wish to identify the
user's partner.
By phrasing the question this starkly, we are able to assess
the extent to which structural information on its own conveys
information about the relationships of interest.

We note that our question has an important contingent nature:
{\em given} that a user has declared a relationship partner, we want
to understand how effectively we can find the partner.
There are different questions that could be asked in a
related vein --- for example, inferring from a user's
network neighborhood whether he or she is in a relationship.
We briefly discuss the connections among these questions 
in a subsequent section, but the problem of identifying 
partners is our main focus here.

As data for our analyses, we principally use
two collections of network neighborhoods from Facebook.
The first consists of the network neighborhoods of approximately 1.3 million
Facebook users, selected uniformly
at random from among all users of age at least 20, with between
50 and 2000 friends, who list a spouse or relationship partner
in their profile.
These neighborhoods have an average of 291 nodes and 6652 links,
for an overall dataset containing
roughly 379 million nodes and 8.6 billion links.

We also employ a smaller dataset --- a sample of approximately 73000
neighborhoods from this first collection, selected uniformly at
random from among all neighborhoods with at most 25000 links.
We refer to this sample as the {\em primary dataset}, and the
larger dataset in the preceding paragraph as the {\em extended dataset}.
We compute our main structural and interaction measures
on both the primary and extended datasets, and these
measures exhibit nearly identical performance on the two datasets.
As we discuss further below, 
we evaluate additional network measures,
as well as more complex combinations of 
measures based on machine learning algorithms, only on the primary dataset.

All Facebook data in these analyses was used anonymously,
and all analysis was done in aggregate.

\section{Embeddedness and Dispersion}

To evaluate approaches for our task of recognizing relationship partners
from network structure, we start with a fundamental baseline ---
the standard characterization of a tie's strength in terms of its
{\em embeddedness}, the number 
of mutual friends shared by its endpoints
\cite{marsden-tie-strength}.
% \cite{marsden-tie-strength, granovetter-weak-ties,granovetter-swt-revisited}.
Embeddedness has also served as the key definition in structural analyses 
for the special case of relationship partners, since it captures how much
the two partners' social circles `overlap'
\cite{felmlee-rel-partner,kalmijn-shared-networks}.
This suggests a natural predictor for identifying a user $u$'s partner:
select the link from $u$ of maximum embeddedness, and propose 
the other end $v$ of this link as $u$'s partner.

We will evaluate this embeddedness-based predictor, and others,
according to their {\em performance}: 
the fraction of instances on which they correctly identify the partner.
Under this measure, embeddedness achieves a performance of $24.7\%$ ---
which both provides evidence about the power of structural information
for this task, but also offers a baseline that other approaches
can potentially exceed.

Next, we show that it is possible
to achieve more than twice the performance of this embeddedness baseline using
our new network measure, {\em dispersion}.
In addition to this relative improvement, 
the performance of our dispersion measure is very high 
in an absolute sense ---
for example, on married users in our sample,
the friend who scores highest under this dispersion
measure is the user's spouse over 60\% of the time.
Since each user in our sample has at least 50 friends, this 
performance is more than 30 times higher than random guessing,
which would produce a performance of at most 2\%.

\begin{figure}[t]
\begin{center}
 \includegraphics[width=0.25\textwidth]{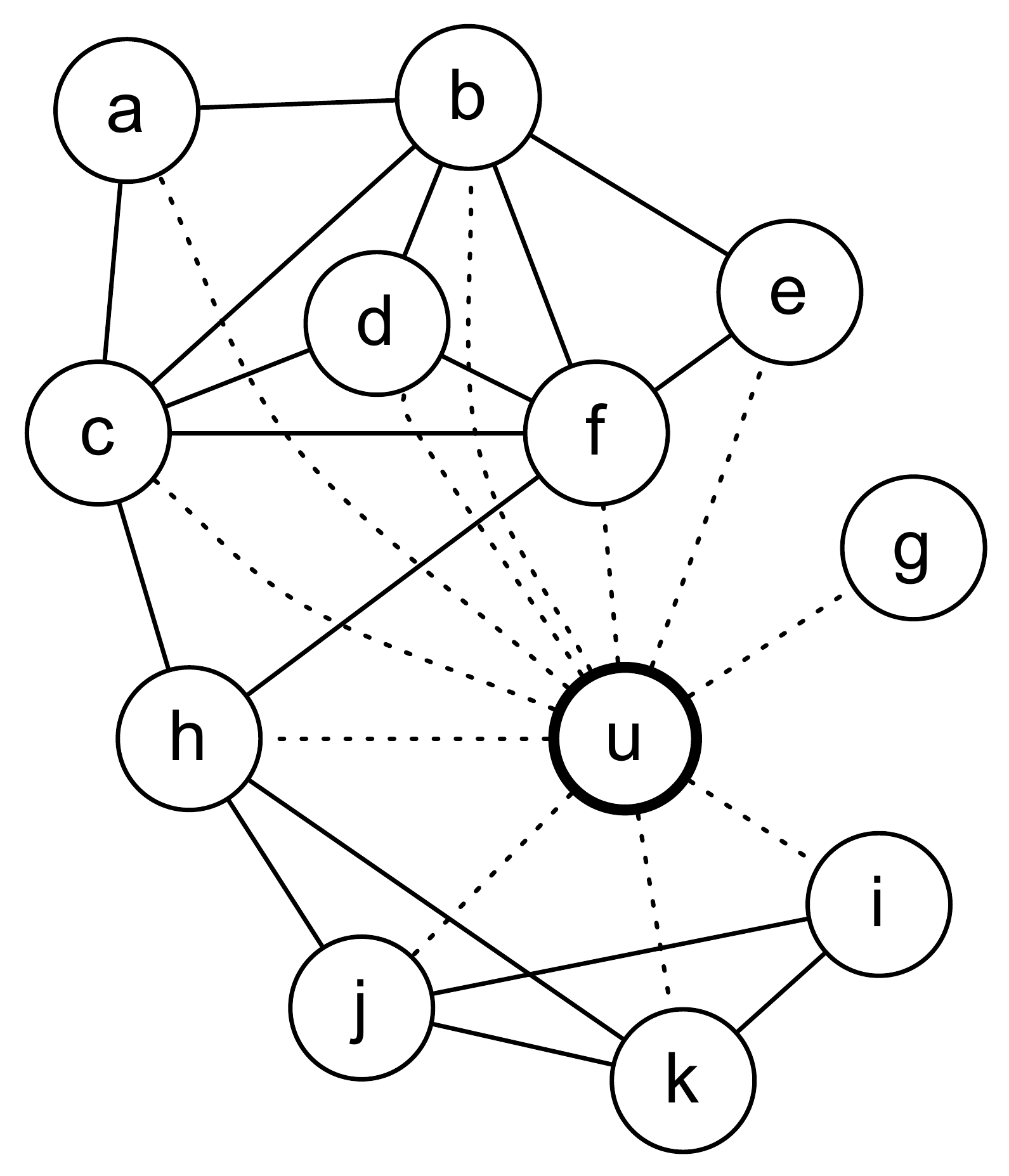}
 \end{center}
\caption{
 \label{fig:nb-synth}
% \footnotesize
A synthetic example
network neighborhood for a user $u$; the links from $u$ to $b$, $c$, and $f$
all have embeddedness 5 (the highest value in this neighborhood),
whereas the link from $u$ to $h$ has an embeddedness of 4.
On the other hand, 
nodes $u$ and $h$ are the unique pair of intermediaries
from the nodes $c$ and $f$ to the nodes $j$ and $k$;
the $u$-$h$ link has greater dispersion
than the links from $u$ to $b$, $c$, and $f$.
\rs 
  }
\end{figure}

\xhdr{Theoretical Basis for Dispersion}
We motivate the dispersion measure by first highlighting a basic 
limitation of embeddedness as a predictor, drawing on the theory
of {\em social foci} \cite{feld-foci}.
Many individuals have large clusters of friends corresponding to well-defined 
foci of interaction in their lives, such as their cluster of
co-workers or the cluster of people with whom they attended college.
Since many people within these clusters know each other, the clusters
contain links of very high embeddedness, even though they do not 
necessarily correspond to particularly strong ties.
In contrast, the links to 
a person's relationship partner or other
closest friends may have lower embeddedness,
but they will often involve mutual neighbors from several
different foci, reflecting the fact that the social orbits 
of these close friends are not bounded within any one focus ---
consider, for example, a husband who knows several of his
wife's co-workers, family members, and former classmates, even though
these people belong to different foci and do not know each other.

Thus, instead of embeddedness, we propose that the link 
between an individual $u$ and his or her partner $v$ should display a
`dispersed' structure: the mutual neighbors of $u$ and $v$ are not
well-connected to one another, and hence $u$ and $v$ act jointly
as the only intermediaries 
between these different parts of the network.
(See Figure \ref{fig:nb-synth} for an illustration.)

We now formulate a sequence of definitions that captures this idea
of dispersion.
To begin, we take the subgraph $G_u$ induced on $u$ and all
neighbors of $u$, and for a
node $v$ in $G_u$ we define $C_{uv}$ to be the set of common neighbors
of $u$ and $v$.
To express the idea that pairs of nodes in $C_{uv}$ should be far
apart in $G_u$ when we do not consider the two-step paths through
$u$ and $v$ themselves, we define the {\em absolute dispersion} 
of the $u$-$v$ link, $disp(u,v)$,
to be the sum of all pairwise distances between nodes in $C_{uv}$,
as measured in $G_u - \{u, v\}$;
that is,
$$disp(u,v) = \sum_{s, t \in C_{uv}} d_v(s,t),$$ where
$d_v$ is a distance function on the nodes of $C_{uv}$.
The function $d_v$ need not be the standard graph-theoretic distance;
different choices of $d_v$ will give rise to different
measures of absolute dispersion.
As we discuss in more detail below, 
among a large class of possible distance functions, we ultimately find the
best performance when we define $d_v(s,t)$ to be the function equal to $1$
when $s$ and $t$ are not directly linked and also
have no common neighbors in $G_u$ other than $u$ and $v$,
and equal to $0$ otherwise.
% do we need "not directly linked"?
For the present discussion, we will use this distance
function as the basis for our measures of dispersion;
below we consider the effect of alternative distance functions.
For example, in Figure \ref{fig:nb-synth}, 
$disp(u,h) = 4$ under this definition and distance function, since 
there are four pairs of nodes in $C_{uh}$ that are not directly
linked and also have no neighbors in common in $G_u - \{u, h\}$.
In contrast, $disp(u,b) = 1$ in Figure \ref{fig:nb-synth}, since $a$ and $e$
form the only pair of non-neighboring nodes in $C_{ub}$ that have no neighbors 
in common in $G_u - \{u, b\}$.

\begin{figure}
\begin{center}
 \includegraphics[width=0.45\textwidth]{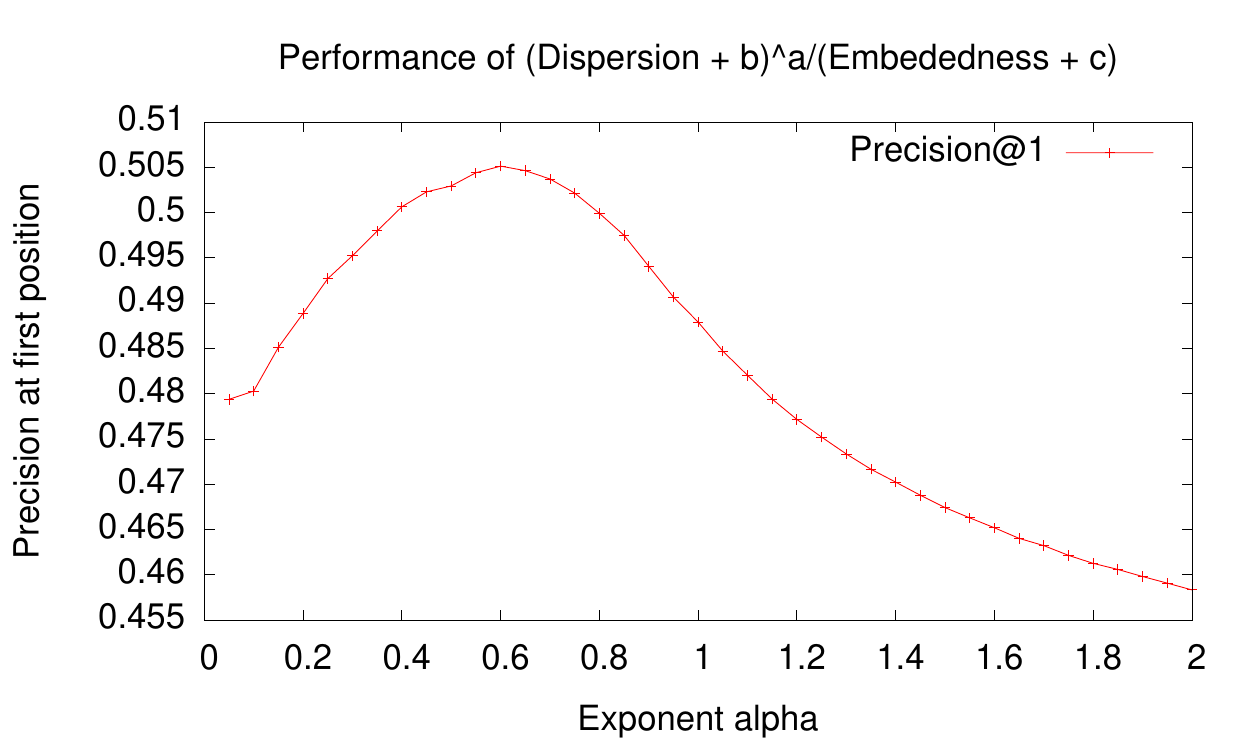}
\end{center}
\caption{
Performance of $(disp(u,v) + b)^\alpha / (emb(u,v) + c)$ as a function of $\alpha$, when choosing optimal values of $b$ and $c$.
  }
\label{fig:perf_by_exp}
\end{figure}

\begin{figure}
\begin{center}
%% perf
\begin{tabular}{|l|c|c|c|c|}
\hline
type & embed & \mbox{rec.disp.} & \mbox{photo} & \mbox{prof.view.}
\\\hline
all & 0.247 & 0.506 & 0.415 & 0.301 \\ \hline 
married & 0.321 & 0.607 & 0.449 & 0.210 \\ \hline 
married (fem) & 0.296 & 0.551 & 0.391 & 0.202 \\ \hline 
married (male) & 0.347 & 0.667 & 0.511 & 0.220 \\ \hline 
engaged & 0.179 & 0.446 & 0.442 & 0.391 \\ \hline 
engaged (fem) & 0.171 & 0.399 & 0.386 & 0.401 \\ \hline 
engaged (male) & 0.185 & 0.490 & 0.495 & 0.381 \\ \hline 
relationship & 0.132 & 0.344 & 0.347 & 0.441 \\ \hline 
relationship (fem) & 0.139 & 0.316 & 0.290 & 0.467 \\ \hline 
relationship (male) & 0.125 & 0.369 & 0.399 & 0.418 \\ \hline 
\end{tabular}
\caption{
\label{table:results}
% \footnotesize
The performance of different measures for identifying
spouses and romantic partners: the numbers in the table give
the {\em precision at the first position} --- the fraction of
instances in which the user ranked first by the measure is
in fact the true partner.
Averaged over all instances, 
recursive dispersion performs approximately twice as well
as the standard notion of embeddedness, and also better overall than
measures based on profile viewing and presence in the same photo.
\rs 
}
\end{center}
\end{figure}

\begin{figure}
\begin{center}
%% perf
\begin{tabular}{|l|c|c|c|c|}
\hline
type & embed & \mbox{rec.disp.} & \mbox{photo} & \mbox{prof.view.}
\\\hline
all & 0.391 & 0.688 & 0.528 & 0.389 \\ \hline 
married & 0.462 & 0.758 & 0.561 & 0.319 \\ \hline 
married (fem) & 0.488 & 0.764 & 0.538 & 0.350 \\ \hline 
married (male) & 0.435 & 0.751 & 0.586 & 0.287 \\ \hline 
engaged & 0.335 & 0.652 & 0.553 & 0.457 \\ \hline 
engaged (fem) & 0.375 & 0.674 & 0.536 & 0.492 \\ \hline 
engaged (male) & 0.296 & 0.630 & 0.568 & 0.424 \\ \hline 
relationship & 0.277 & 0.572 & 0.460 & 0.498 \\ \hline 
relationship (fem) & 0.318 & 0.600 & 0.440 & 0.545 \\ \hline 
relationship (male) & 0.239 & 0.546 & 0.479 & 0.455 \\ \hline 
\end{tabular}
\caption{
\label{table:family}
% \footnotesize
The performance of the four measures from Figure \ref{table:results}
when the goal is to identify the partner or a family member in the
first position of the ranked list.
The difference in performance between the genders has largely
vanished, and in some cases is inverted relative to Figure \ref{table:results}.
\rs 
}
\end{center}
\end{figure}

\xhdr{Strengthenings of Dispersion}
We can learn a function 
that predicts whether or not
$v$ is the partner of $u$ in terms of the two variables
$disp(u,v)$ and $emb(u,v)$, where the latter denotes the embeddedness
of the $u$-$v$ link.  
We find that performance is highest for functions that are monotonically
increasing in $disp(u,v)$ and monotonically decreasing in $emb(u,v)$:
for a fixed value of $disp(u,v)$, increased embeddedness is in fact
a negative predictor of whether $v$ is the partner of $u$.
A simple combination of these two quantities that comes within
a few percent of more complicated functional forms can be obtained
by the expression
$disp(u,v) / emb(u,v)$, which we term the {\em normalized dispersion}
$norm(u,v)$ since it normalizes the absolute dispersion by the embeddedness.
Predicting $u$'s partner to be the individual
$v$ maximizing $norm(u,v)$ gives 
the correct answer in $48.0\%$ of all instances.

% calc \(22422*.34056+42170*.60242+7701*.45358\)/\(22422+42170+7701\)
% .50534778332618649108
There are two strengthenings of the normalized dispersion that
lead to increased performance.
The first is to rank nodes by a function of the form
$(disp(u,v) + b)^\alpha / (emb(u,v) + c)$.
Searching over choices of $\alpha$, $b$, and $c$ leads to 
maximum performance of $50.5\%$ at $\alpha = 0.61$, $b = 0$, and $c = 5$;
see Figure \ref{fig:perf_by_exp}.
Alternately, one can strengthen performance by applying the idea of 
dispersion recursively --- identifying nodes $v$
for which the $u$-$v$ link achieves a high normalized dispersion based on
a set of common neighbors $C_{uv}$ who, in turn, also have
high normalized dispersion in their links with $u$.
To carry out this recursive idea, we assign values to the nodes
reflecting the dispersion of their links with $u$, and then update these values
in terms of the dispersion values associated with other nodes.
Specifically, we initially define $x_v = 1$ for all neighbors $v$ of $u$,
and then iteratively update each $x_v$ to be 
$$\frac{    \sum_{w \in C_{uv}} x_w^2 
        + 2 \sum_{s, t \in C_{uv}} d_v(s,t) x_s x_t}
       {emb(u,v)}.$$
Note that after the first iteration, $x_v$ is $1 + 2 \cdot norm(u,v)$,
and hence ranking nodes by $x_v$ after the first iteration is equivalent
to ranking nodes by $norm(u,v)$.
We find the highest performance when we rank nodes by the values of
$x_v$ after the third iteration. 
% the top-ranked node $v$ is the partner
% of $u$ in $50.6\%$ of all instances, more than twice the performance of
% ranking based on embeddedness, and more than a 25-fold improvement
% over the baseline of random selection.
For purposes of later discussion, we will call this value $x_v$
in the third iteration the {\em recursive dispersion} $rec(u,v)$,
and will focus on this as the main examplar from our family
of related dispersion-based measures.
(See the Appendix for further mathematical properties of 
the recursive dispersion.)

\section{Performance of Structural and Interaction Measures}

We summarize the performance of our methods 
in Figure \ref{table:results}.
Looking initially at just the 
first two columns in the top row of numbers (`all'),
we have the overall performance of
embeddedness and recursive dispersion --- the fraction of instances
on which the highest-ranked node under these measures is in 
fact the partner.  
As we will see below in the discussion around 
Figure \ref{tab:full-dispersion},
recursive dispersion also has higher performance than
a wide range of other basic structural measures.

We can also compare these structural measures to features derived
from a variety of different forms of real-time interaction between users --- 
including the viewing of profiles, sending of messages, 
and co-presence at events.
The use of such `interaction features' as a comparison baseline is
motivated by the way in which tie strength can be estimated from 
the volume of interaction between two 
people
\cite{eagle-inferring-friendship,jones-tie-strength}.
% \cite{eagle-inferring-friendship,jones-tie-strength,gilbert-tie-strength}.
Within this category of interaction features,
the two that consistently display
the best performance are to rank 
neighbors of $u$ by the number of photos in which they appear with $u$,
and to rank neighbors of $u$ by the total number of times that
$u$ has viewed their profile page in the previous 90 days.
The last two columns of Figure \ref{table:results} show
the performance of these two measures;
on the set of instances as a whole, recursive dispersion performs
better than these features.

The remaining rows of Figure \ref{table:results}
show the performance of these measures on different subsets
of the data.
Most users who report a relationship partner on Facebook list themselves
as either `married' or `in a relationship,' with a smaller number who are
`engaged.'
The performance of the structural measures is much higher for
married users (60.7\%) than for users in a relationship (34.4\%);
the opposite is true for profile viewing, which in fact achieves
higher performance than recursive dispersion for users
in a relationship.
The performance for users who are engaged is positioned
between the extremes of `married' and `in a relationship.'

In addition, we see important differences based on gender.
The performance of structural measures is significantly higher
for males than for females, suggesting some of the ways
in which relationship partners produce more visible
structural effects --- at least according to these measures ---
on the network neighborhoods of men.
And for certain more focused
subsets of the data, the performance is even stronger;
for example, on the subset corresponding to
married male Facebook users in the United States,
the friend with the highest recursive dispersion is the user's
spouse $76.9\%$ of the time.

We can also evaluate performance on the subset of users in same-sex
relationships.  Here we focus on users whose status is
`in a relationship.'\footnote{There is significant informal
evidence that the `married' relationship status is employed
by younger users of the same gender for a range of purposes even
when they are not, in fact, married.  While we only include users
of age at least 20 in our sample, the effect is present in that age range.
Listing a relationship status
that does not correspond to one's off-line relationship is of course
a concern across all categories of users, but from investigation of this
issue, the `married' status for users
of the same gender is the only category for which we see evidence
that this is a significant factor.}
The relative performance of our structural measures is exactly
the same for same-sex relationships as for the set of
all relationships, with recursive dispersion achieving close
to twice the performance of embeddedness, and slightly higher performance
than absolute and normalized dispersion.
For female users, the absolute level of performance is almost
identical regardless of whether their listed partner is female or male;
for male users, the performance is significantly higher for
relationships in which the partner is male.
(For a same-sex relationship listed by a male user,
recursive dispersion identifies the partner with a performance
of $.450$, in contrast with the performance of $.369$ for all
partners of male users shown in Figure \ref{table:results}.)

Finally, returning to the set of all relationships,
when the user $v$ who scores highest under one of these measures
is not the partner of $u$, what role does $v$ play among $u$'s
network neighbors?
We find that $v$ is often a family member of $u$;
for married users (Figure \ref{table:family}), the friend $v$ that
maximizes $rec(u,v)$ is the partner or a family member over $75\%$
of the time.
We also see that when we ask for the top-ranked friend to be
either the partner or a family member, rather than just the partner,
the performance gap between the genders essentially vanishes
in the case of married users, and becomes inverted in the case
of users in a relationship --- in this latter case, female users
are more likely to have their partner or a family member at the top
of the ranking by recursive dispersion.

\begin{figure}
\begin{center}
\begin{tabular}{|l|l|c|c|c|c|}
\hline
distance & type & all & marr. & eng. & rel. \\ \hline \hline
threshold 2 & absolute & 0.279 & 0.361 & 0.205 & 0.152 \\ \hline 
  & normalized & 0.305 & 0.394 & 0.227 & 0.168 \\ \hline 
  & recursive & 0.210 & 0.279 & 0.141 & 0.105 \\ \hline 
\hline
threshold 3 & absolute & 0.430 & 0.530 & 0.359 & 0.270 \\ \hline 
  & normalized & 0.486 & 0.588 & 0.425 & 0.322 \\ \hline 
  & recursive & 0.506 & 0.607 & 0.446 & 0.344 \\ \hline 
\hline
threshold 4 & absolute & 0.473 & 0.568 & 0.414 & 0.321 \\ \hline 
  & normalized & 0.483 & 0.570 & 0.434 & 0.342 \\ \hline 
  & recursive & 0.455 & 0.539 & 0.405 & 0.319 \\ \hline 
\hline
diff component & absolute & 0.380 & 0.461 & 0.317 & 0.253 \\ \hline 
  & normalized & 0.364 & 0.433 & 0.308 & 0.258 \\ \hline 
  & recursive & 0.323 & 0.384 & 0.276 & 0.228 \\ \hline 
\hline
diff community & absolute & 0.286 & 0.368 & 0.212 & 0.160 \\ \hline 
  & normalized & 0.296 & 0.379 & 0.221 & 0.167 \\ \hline 
  & recursive & 0.216 & 0.283 & 0.156 & 0.115 \\ \hline 
\hline
spring length & absolute & 0.379 & 0.474 & 0.307 & 0.229 \\ \hline 
  & normalized & 0.454 & 0.553 & 0.387 & 0.296 \\ \hline 
  & recursive & 0.396 & 0.480 & 0.341 & 0.261 \\ \hline 
\hline
\end{tabular}
\end{center}
\caption{Performance of variants of the dispersion measure using different underlying distance functions.  \label{tab:full-dispersion}}
\end{figure}

% betw: .441 .535 .374 .293
% burt: .307 .394 .232 .171

\begin{figure}
\begin{center}
\begin{tabular}{|l|l|c|c|c|c|}
\hline
measure & all & married & engaged & relationship \\ \hline \hline
betweenness & 0.441 & 0.535 & 0.374 & 0.293 \\ \hline \hline
network constraint & 0.307 & 0.394 & 0.232 & 0.171 \\
\hline
\end{tabular}
\end{center}
\caption{Performance of betweenness and network constraint as alternate
measures of bridging.  \label{tab:betw-nc}}
\vspace*{-0.15in}
\end{figure}

\xhdr{A Broader Set of Measures}
Since measures of dispersion are based on an underlying 
distance function $d_v$, it is interesting to investigate
how the performance depends on the choice of $d_v$.
In Figure \ref{tab:full-dispersion}, we consider
dispersion measures based on a range of natural choices for $d_v$.
as follows.

\begin{itemize}
\item First, we can set a distance threshold $r$, and declare
that $d_v(s,t) = 1$ when $s$ and $t$ are at least $r$ hops apart
in $G_u - \{u, v\}$, and $d_v(s,t) = 0$ otherwise.
The measure of dispersion we use above corresponds 
to the choice $r = 3$; setting the threshold $r = 2$ simply requires
that $s$ and $t$ are not directly connected, while setting $r = 4$
imposes a stricter requirement.
\item In a related vein, we could declare $d_v(s,t) = 1$ if $s$ and $t$
belong to different connected components of $G_u - \{u, v\}$,
and $d_v(s,t) = 0$ otherwise.
This in effect follows the preceding approach, but with the 
distance threshold $r$ conceptually set to be infinite.
\item Since the idea of
dispersion at a more general level is based on the notion that the 
common neighbors of $u$ and $v$ should belong to different `parts'
of the network, it is also natural to divide $G_u - \{u\}$ into
{\em communities} according to a network clustering heuristic, 
and then declare $d_v(s,t) = 1$ if and only if
$s$ and $t$ belong to different communities.
For this purpose, we use the {\em Louvain method}
of Blondel et al
for optimizing
modularity
\cite{blondel-louvain-method},
% \cite{blondel-louvain-method,clauset-modularity-opt}, 
as implemented in the software package NetworkX.
There is a wide range of methods available for inferring
communities from network data \cite{newman-sirev};
we choose the Louvain method as a baseline because the 
graph $G - \{u\}$ for most users
tends to have clearly defined {\em modules} of nodes,
corresponding roughly to social foci, with a high density of links
inside each module and a low density of links between them.
Such modular structure is what the Louvain method is designed to identify.
\item Related to partitions into communities, one can embed 
$G_u -\{u\}$ in the plane using an energy-minimization heuristic that
treats each link of the graph as a spring with a fixed rest length.
After computing such a spring embedding of $G_u - \{u\}$, one can then
define $d_v(s,t)$ simply to be the distance between the locations
of $s$ and $t$ in their embedding in the plane.
Here too we use a heuristic implemented in NetworkX, in this
case for spring embedding.
\end{itemize}

Figure \ref{tab:full-dispersion} shows the performance of 
the absolute, normalized, and recursive dispersion based on
all these possible distance functions $d_v$.
(For all these distance functions, when the recursive process
was carried out, the iteration other than the first producing the highest
performance was always the third iteration, and so we continue to use the
results in this third iteration as the definition of recursive dispersion.)
As noted above, recursive dispersion using
a distance function $d_v$ based on a distance threshold of $3$
produces the highest accuracy.  

Finally, there are other measures of bridging that cannot
be naturally expressed using the framework of dispersion.
Two standard such measures are 
are {\em betweenness} and {\em network constraint} 
\cite{burt-struct-holes-book}
applied to $G_u - \{u\}$.
Figure \ref{tab:betw-nc}
shows their performance;
both score below the strongest dispersion measures,
although betweenness has strong performance.

\begin{figure}
\begin{center}
\includegraphics[width=0.41\textwidth]{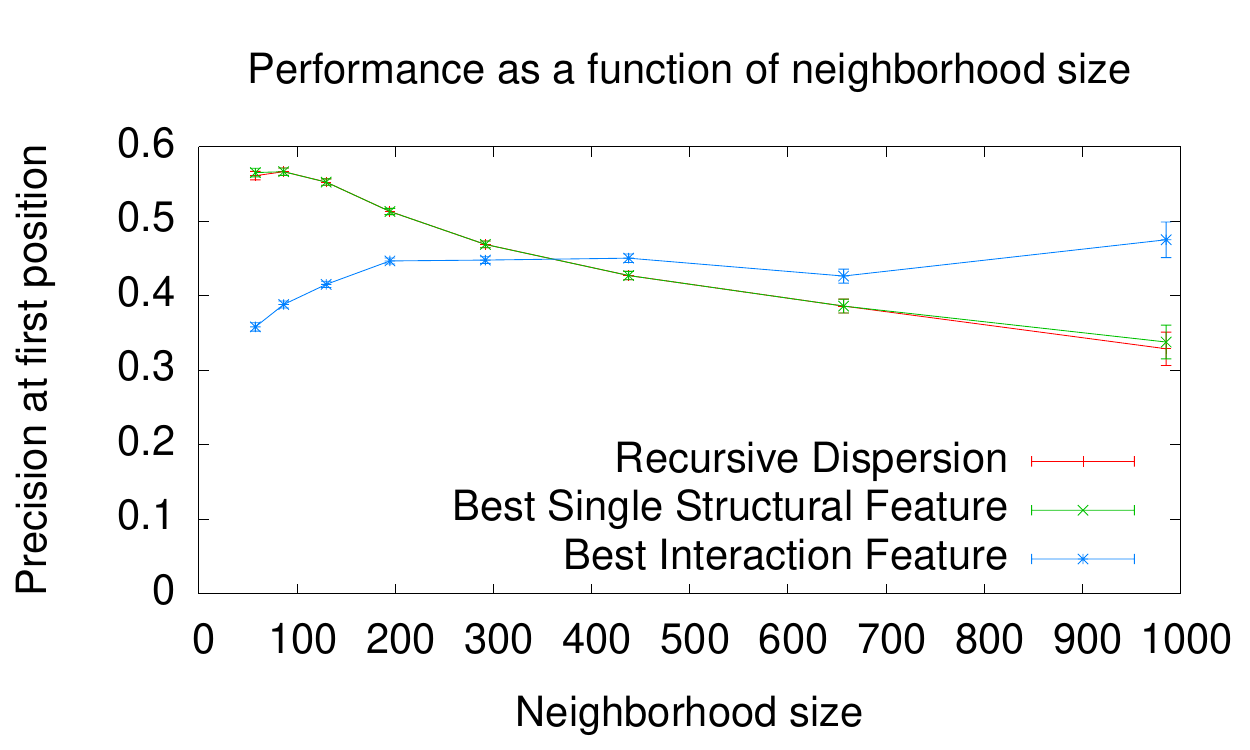}
\end{center}
\caption{
\label{fig:perf_by_deg}
Performance as a function of user $u$'s neighborhood size.
  }
\vspace*{-0.15in}
\end{figure}

\begin{figure}
\begin{center}
\includegraphics[width=0.32\textwidth]{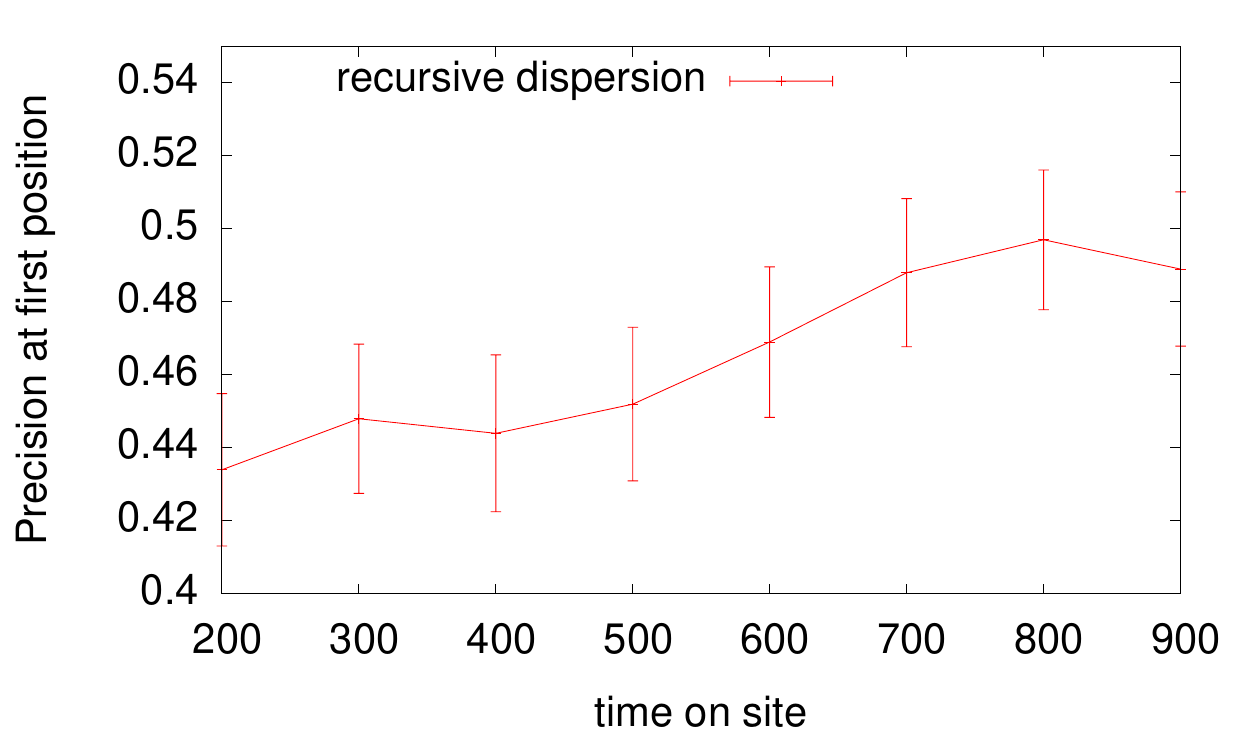}
\end{center}
\caption{
\label{fig:plot-fbage01}
Performance as a function of time on site, for a subset of
users where we control both the neighborhood size (between 100 and 150)
and the time since the relationship was reported 
(between 100 and 200 days).
\rs \rs \rs \rs 
  }
\end{figure}

\xhdr{Performance as a Function of Neighborhood Size and Time on Site}
A further important source of variation among users is in the 
size of their network neighborhoods and the amount of time since
they joined Facebook.
These two properties are related; after a user joins
the site, his or her network neighborhood will generally grow
monontonically over time.
This growth has two potential effects on the level of performance 
for our problem, acting in opposite directions: 
a more mature network neighborhood will have a greater level of complexity,
which may hurt performance; but it will also likely reflect 
the user's off-line relationships in richer detail, which may help
performance.
It is thus natural to evaluate performance as a function of these
parameters.

We begin by considering the size of the network neighborhood of
the user $u$;
recall that in our data, this size ranges from 50 to 2000.
We find that recursive dispersion performs well across this full range:
Figure \ref{fig:perf_by_deg} shows that, 
while performance is best when the neighborhood size is
around 100 nodes (56\%), it only drops moderately (to 33\%)
as the neighborhood size increases by an order of magnitude to 1000 nodes.
This decline is quite small compared to the decline of the
baseline that simply guesses a node uniformly at random, which would
decrease in performance by a factor of $10$ (from $1/100$ to $1/1000$).
The modest decline of recursive dispersion may reflect the ways
in which larger neighborhoods are also more informative about 
a user's full set of off-line relationships, which helps offset the
considerably increased number of options for identifying the
relationship partner.
Furthermore, recursive dispersion 
is the highest-performing structural measure among those considered 
for every range of neighborhood sizes except the extremes
(50 to 100, and above 1000); at the extremes, the (non-recursive)
normalized dispersion is slightly better, 
although even here
the recursive measure is within the margin of error of the best.
Note that the median neighborhood size in our dataset is 205.

The benefits of large neighborhoods are reflected even more clearly
when we consider the performance of interaction features --- their
performance tends to be approximately constant, or even increasing,
as a function of neighborhood size.
To elaborate on the arguments discussed above for how a larger
neighborhood may help performance, we make two further observations.
First, users with large neighborhoods also tend to be more
active, and thus the relative variance of the interaction signals is smaller.
Second, despite their large neighborhoods, 
previous analysis \cite{marlow-facebook-tie-strength-ext}
has shown that the number of relationships that are actively maintained
grows slowly in the total neighborhood size,
and so the number of plausible candidates for the relationship
partner grows more slowly than the pure neighborhood size would suggest.  

\begin{figure}
\begin{center}
%% perf
\begin{tabular}{|l|c|c|c|c|c|}
\hline
type & max. & max. & all. & all. & comb. \\
     & struct. & inter. & struct. & inter. & 
\\\hline
all & 0.506 & 0.415 & 0.531 & 0.560 & 0.705
\\\hline
married & 0.607 & 0.449 & 0.624 & 0.526 & 0.716
\\\hline
engaged & 0.446 & 0.442 & 0.472 & 0.615 & 0.708
\\\hline
relationship & 0.344 & 0.441 & 0.377 & 0.605 & 0.682
\\\hline
\end{tabular}
\caption{
\label{table:ml_perf}
% \footnotesize
The performance of methods based on machine learning that combine
sets of features.
The first two columns show the highest performing individual structural
and interaction features; the third and fourth columns show the
highest performance of machine learning classifiers that combine
structural and interaction features respectively; and the fifth
column shows the performance of a classifier that combines all
structural and interaction features together.
\rs 
}
\end{center}
\vspace*{-0.12in}
\end{figure}

\begin{figure*}[t]
\begin{center}
% \hspace*{-0.1\textwidth}
% \subfigure[]{
 \includegraphics[width=0.45\textwidth]{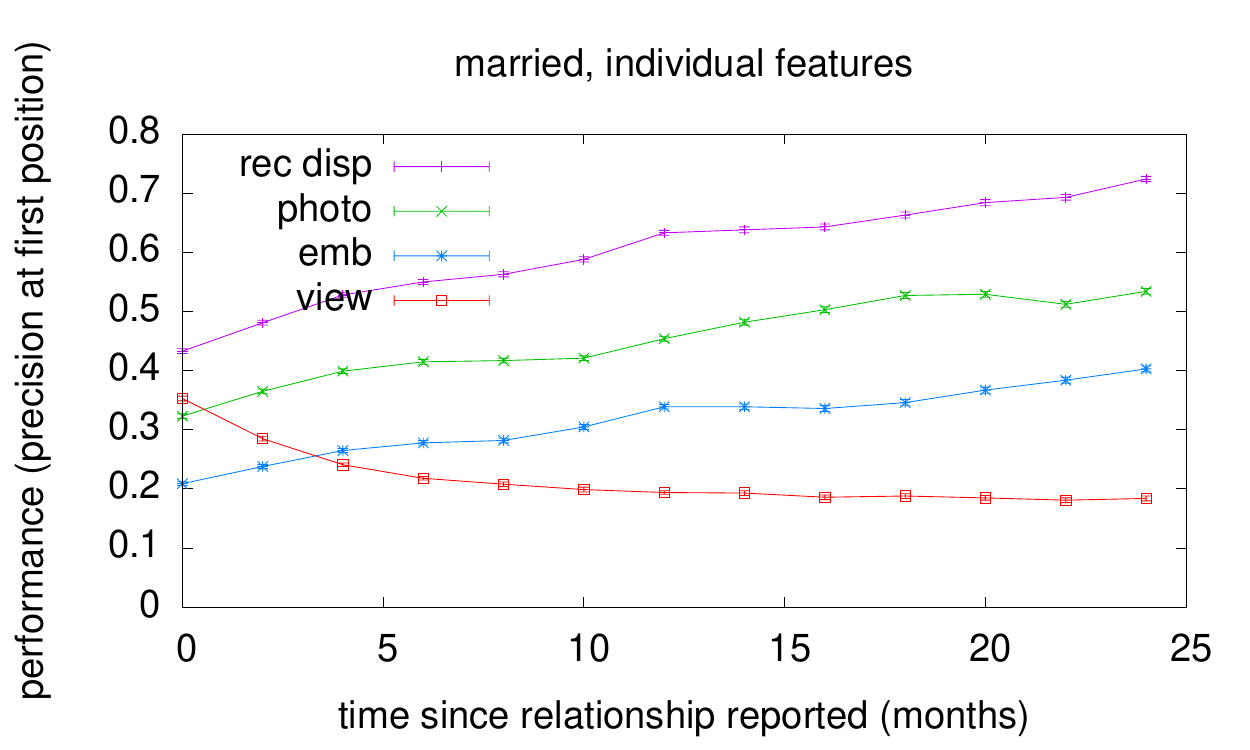}
 \label{fig:age-m}
% }
% \subfigure[]{
 \includegraphics[width=0.45\textwidth]{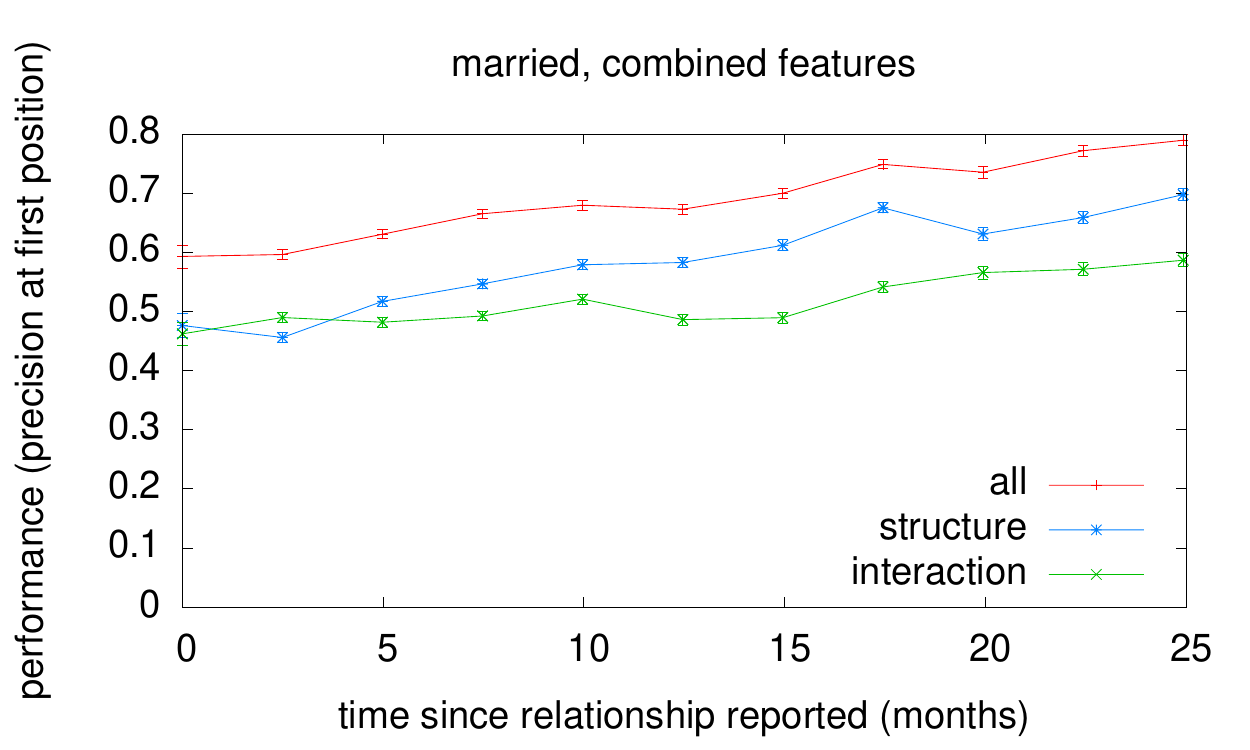}
 \label{fig:perf-married}
% }
% \hspace*{-0.1\textwidth}
% \subfigure[]{
 \includegraphics[width=0.45\textwidth]{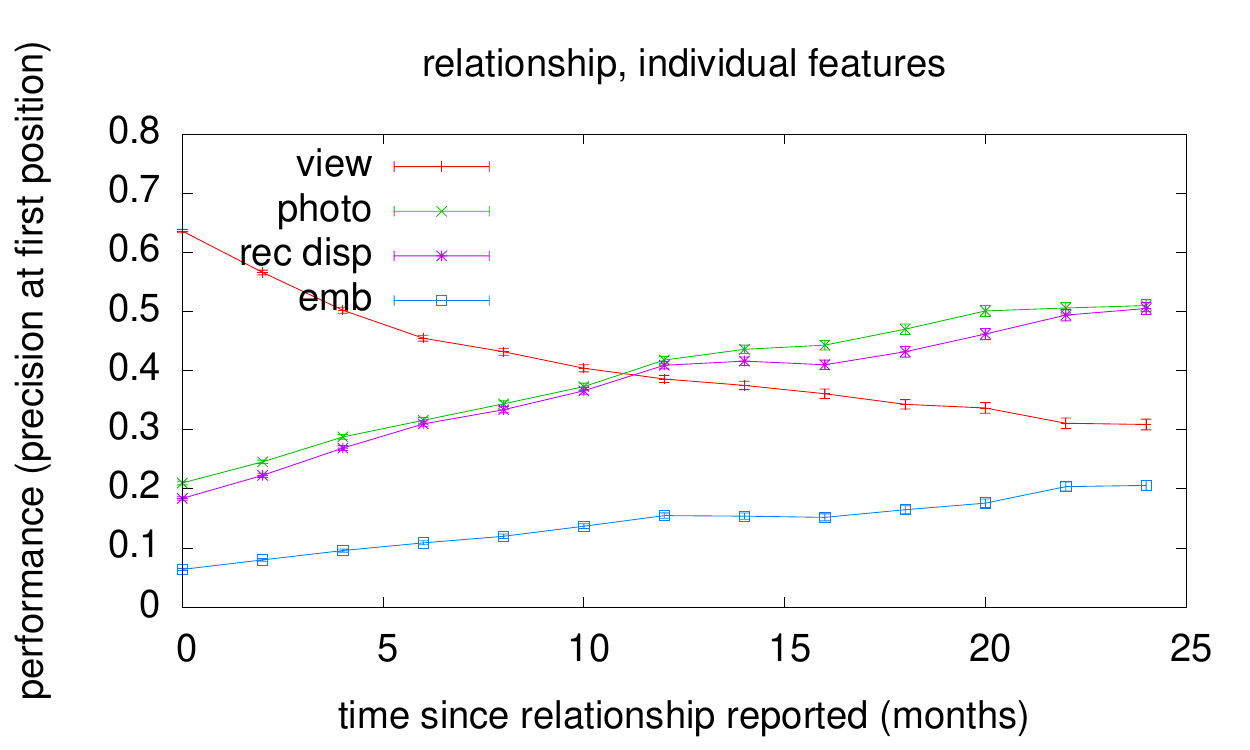}
 \label{fig:age-r}
% }
% \subfigure[]{
 \includegraphics[width=0.45\textwidth]{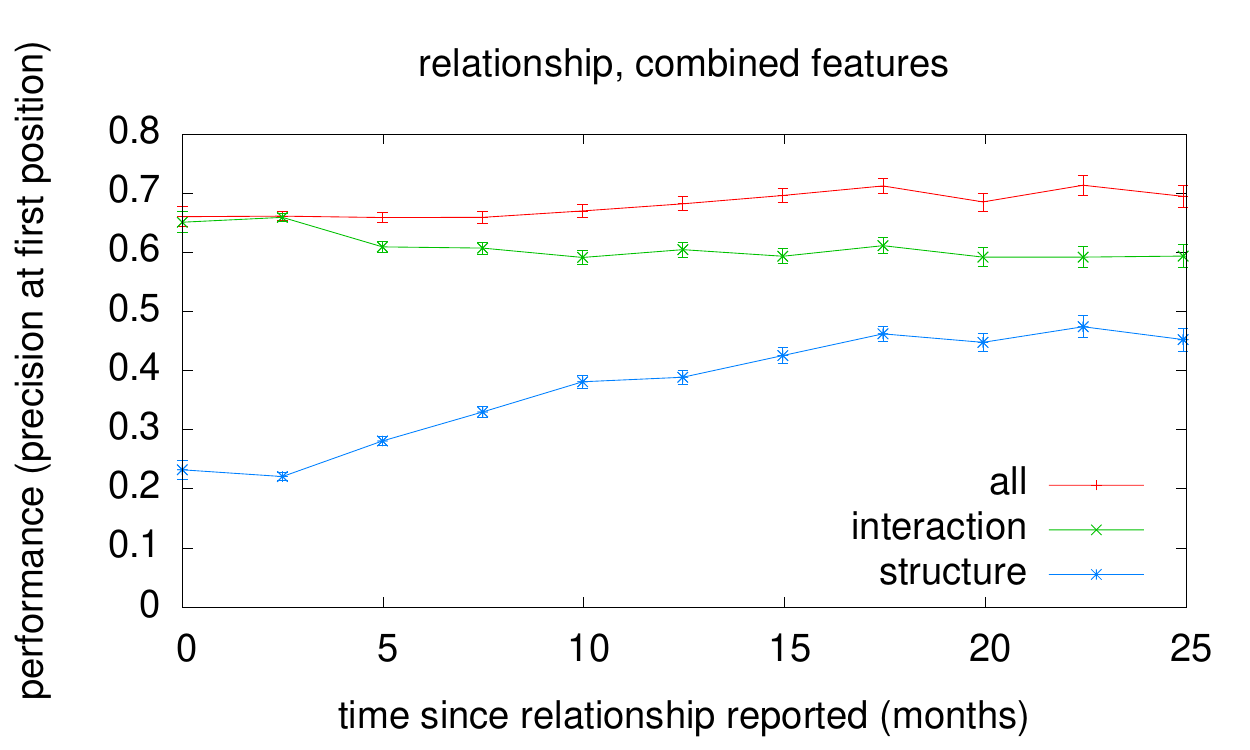}
 \label{fig:perf-relationship}
% }
\end{center}
\caption{
% \footnotesize
Performance at identifying partners 
for users who are (top row) married and
(bottom row) in a relationship, 
as a function of the time since the relationship was reported on Facebook.
In the two figures in the left-hand column, 
the four measures from Figure \ref{table:results}
are used individually.
Here the structural measures have higher
performance on older relationships, while the profile viewing feature 
has lower performance on older relationships;
the profile viewing feature outperforms recursive dispersion for
relationships reported recently, with the relative performance
crossing at approximately one year.
In the two figures in the right-hand column, 
the performance as a function of time is shown for prediction rules
constructed using machine learning on a large set of features.
Separate curves show the performance using only 
structural features, only interaction features, 
and when using the union of the
two.  Structural features peform best on older marriages, 
while interaction features perform best on new relationships.
\label{fig:age}
\rs
  }
\end{figure*}

We also consider a user's {\em time on site} --- the number of days
since they joined Facebook.
This is strongly correlated with neighborhood size, since users
continue acquiring friendship links over their time on Facebook,
and it is also correlated with the time since the relationship 
was first reported. (As we will see later in 
Figure \ref{fig:age}, performance varies as a function of this
latter quantity as well.)
To understand whether there is any effect of a user's time on site
beyond its relation to these other parameters, we consider a subset
of users where we restrict the neighborhood size to lie 
between 100 and 150, and the time since the relationship was reported
to lie between 100 and 200 days.
Figure \ref{fig:plot-fbage01} shows that for this subset, there
is a weak increase in performance as a function of time on site;
while the effect is not strong, 
it points to a further source of enhanced performance
for users with mature neighborhoods.

\section{Combining Features using Machine Learning}

Different features may capture different aspects of the user's
neighborhood, and so it is natural to ask how well we can predict partners when
combining information from many structural or interaction features 
via machine learning.

\xhdr{Machine Learning Techniques}
For our machine learning experiments, we compute $48$ structural features and
$72$ interaction features for all of the nodes in the neighborhoods
from our primary dataset.
% (consisting of approximately 73000 neighborhoods).
This gives us a total of approximately 18.7 million labeled instances 
with $120$ features each ---
each instance consists of a node $v$ in a neighborhood $G_u$,
with a positive label indicating $v$ is the partner
of $u$, or a negative label indicating $v$ is not.

The 48 structural features are
the absolute and normalized dispersion
based on six distinct distance functions defined 
for Figure \ref{tab:full-dispersion},
as well as the recursive versions using iterations 2 through 7
(recall that the recursive dispersion 
corresponds to the third iteration, and is hence included).
The 72 interaction features represent a broad
range of properties including the number of photos in which $u$ and $v$ 
are jointly tagged, the number of times $u$ has viewed $v$'s profile
over the last 30, 60, and 90 days, the number of messages sent from $u$ to $v$,
the number of times that $u$ has `liked' $v$'s content and vice versa,
and measures based on a number of forms of interaction closely related 
to these.

To improve
the performance of the learning algorithms, we transformed each of the $120$
features into $4$ different versions: (a) the raw feature, (b) a normalized
version of the feature with mean 0 and standard deviation 1, 
(c) a rank version of
the feature (where the 
individual with highest score on this feature has rank 1, 
and other individuals are sorted in ascending rank order from there),
and (d) a rank-normalized version where we divide (c) by total number of
friends a user has.  Thus, the input to our machine learning algorithms has
$480$ features derived from $120$ values per instance.
In addition to the full set of features, we also compute performance 
using only the structural features, and only the interaction features.

We performed initial experiments with different machine learning algorithms
and found that gradient tree boosting
\cite{friedman-gradient-boosting}
out-performed logistic regression, as
well as other tree-based methods.  Thus, all of our in-depth analysis is
conducted with this algorithm.
In our experiments, we divide the data so that 50\% of the users go into
a training set and 50\% go into a test set.  We perform 12 such
divisions into sets A and B; for each division we train on set A
and test on B, and then train on B and test on A.  
For each user $u$, we average
over the 12 runs in which $u$ was a test user to get a final prediction.

\xhdr{Performance of Machine Learning Methods}
We find (Figure
\ref{table:ml_perf}) that by using boosted decision trees to combine all of the
$48$ structural features we analyzed, we can increase performance from
$50.8\%$ to $53.1\%$.  
We can use the same technique to predict relationships
based on interaction features.  We find that, overall, interaction
features perform slightly better than structural features (56.0\% vs. 53.1\%),
though for married users, structural features do much 
better (62.4\% vs. 52.6\%).  
In addition, on all categories we find that the combination of
interaction features and structural features significantly outperforms either
on its own.  When combining all features with boosted trees, the top
predicted friend is the user's partner 70.5\% of the time.

%      single all    single all   all
%      struct struct coeff  coeff
% 2 => 34.3   37.7          60.5  68.2
% 4 => 60.6   62.4          52.6  71.6
% 5 => 44.5   47.2          61.5  70.8
% all=>50.8   53.1          56.0  70.5

\xhdr{Machine Learning to Predict Relationship Status}
Earlier we noted that our focus is on the problem of identifying
relationship partners for users where we know that they are in
a relationship.
It is natural to ask about the connection to a related but distinct
question --- estimating whether an arbitrary user is in a relationship or not.

This latter question is quite a different issue, and it seems likely to
be more challenging and to require a different set of techniques.
To see why, consider a user $u$ who has a link of high dispersion to
a user $v$.
If we know that $u$ is in a relationship, then $v$ is a good candidate
to be the partner.
But our point from the outset has been that methods based on 
dispersion are useful more generally to identify individuals $v$
with interesting connections to $u$, in the sense that they have
been introduced into multiple foci that $u$ belongs to.
A user $u$ can and generally will have such friends even
when $u$ is not in a romantic relationship.
For example, Figure \ref{table:family} suggests that family members
often have this property, and this can apply to users who are
not in romantic relationships as well as to users in such relationships.
Thus, simply knowing that $u$ has links of high dispersion 
should not necessarily give us much leverage in estimating
whether $u$ is in a relationship.

We now describe some basic machine-learning results that bear out
this intuition.  We took approximately 129,000 Facebook users,
sampled uniformly over all users of age at least 20
with between 50 and 2000 friends.
40\% of these users were single, while the remaining were either in a
relationship, engaged, or married.  
We attempt two different prediction tasks: first, determining whether
a user is in any sort of a relationship; and second, an easier
task in which we look only at single and married users, and
attempt to determine which category a user belongs to.
We consider three different sets of features for these tasks:
(i) demographic features (age, gender, country, and time on site);
(ii) structural features of the network neighborhood, 
based on the definitions presented earlier; and (iii) the union
of these two sets.

Figure \ref{table:predict_relationship} shows the performance on these 
tasks.  Because age is a 
powerful feature for predicting relationship status, demographic
features do well.
Network features are not as strong, reflecting the notion discussed
above that even users not in relationships have friends with
similar structural properties.
Despite this, network features convey non-trivial 
information about relationship status; they perform significantly
above baseline prediction on their own, and 
add predictive power to demographic features.  Of
the network features, the maximum normalized dispersion 
(where the maximum is taken over all of the user's friends) 
has the highest feature importance.

\begin{figure}
\begin{center}
%% perf
\begin{tabular}{|l|c|c|c|c|}
\hline
Task & baseline & demo. & network & both
\\\hline
Single vs. Any Rel. & 59.8\% & 67.9\% & 61.6\% & 68.3\%  \\ \hline 
Single vs. Married & 56.6\% & 78.0\% & 66.1\% & 79.0\%  \\ \hline 
\end{tabular}
\caption{
\label{table:predict_relationship}
% \footnotesize
Performance on predicting relationship status.
Baseline accuracy comes from predicting the more common class in each of
the classification tasks.  Demographic features seem more important,
but network features also provide incremental improvement.
\rs \rs 
}
\end{center}
\end{figure}

%single vs any relationship:
%baseline: 59.76%
%trained-all: 68.28%
%trained-network: 61.56%
%trained-demo: 67.88%

%single vs married
%baseline: 56.56%
%trained-all: 79.03%
%trained-network: 66.14%
%trained-demo: 77.97%

\section{Temporal Properties}

We now explore some of the ways in which these measures change over
time.
% \cite{le-rel-dissolution}.
% \cite{kalmijn-shared-networks,kim-stiff-close-rel,le-rel-dissolution}.
We first consider how performance varies
based on the time since the relationship was first reported
by the user --- 
an approximate surrogate for the age of the relationship itself,
although the relationship may clearly have existed for some
time before it was reported (especially in the case of users
who are already in a relationship when they join Facebook).
We find (Fig.~\ref{fig:age}) that the structural measures are more
accurate on older relationships than on newer ones, 
while the profile viewing feature is
less accurate; in effect, the structural signature of the relationship
needs time to `burn in' to the network, while the interaction 
level via profile viewing is high almost immediately.
For married users, recursive dispersion has the highest
performance across the full time range, but for users 
in a relationship, an interesting crossover occurs: for relationships
less than a year old, the profile viewing featured produces the
highest performance, and then recursive dispersion and photo viewing
surpass it at approximately one year.
This illustrates a trade-off between a decreasing level of 
observation as a relationship goes on, contrasted with an 
increasing level of dispersion in the network 
as the link structure adapts around the two individuals.

\begin{figure}[t]
\begin{center}
% \hspace*{-0.1\textwidth}
 \includegraphics[width=0.45\textwidth]{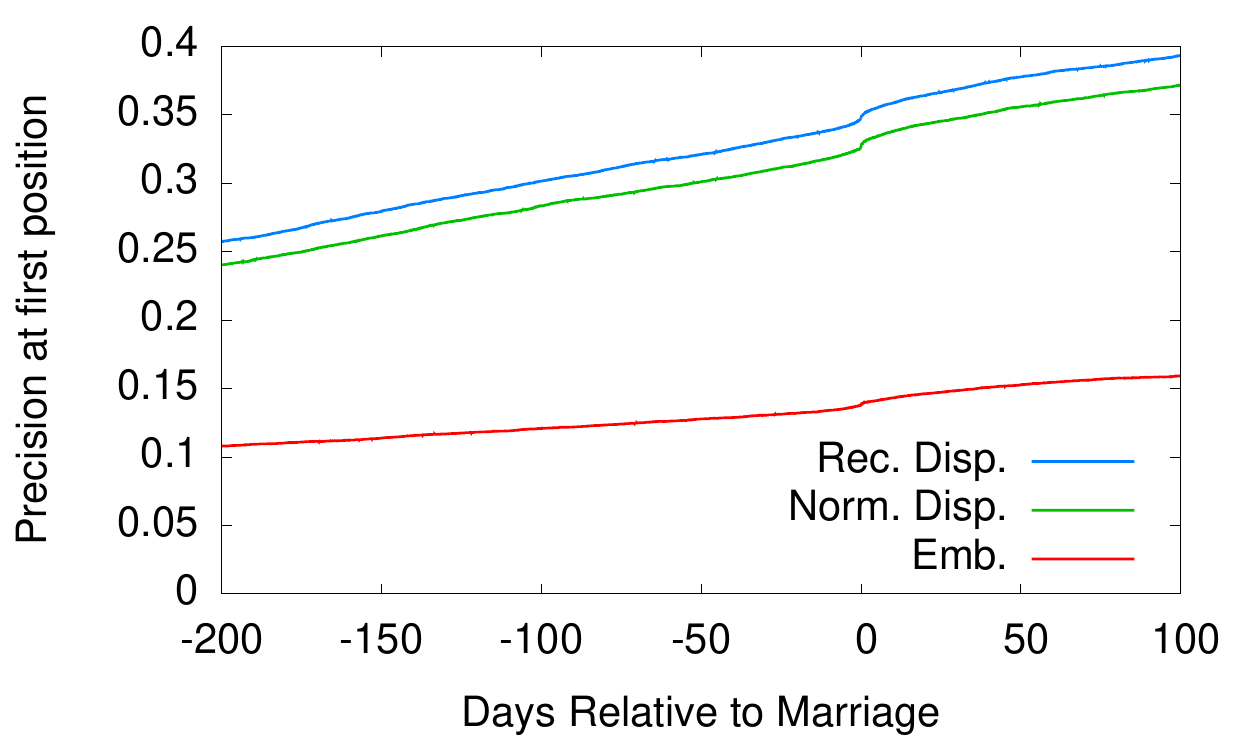}
\caption{
 \label{fig:over-time-perf}
% \footnotesize
The performance of embeddedness and normalized
and recursive dispersion increase as two people approach
the point at which they announce their relationship status as `married'
(at time 0).
\rs  
}
\end{center}
\end{figure}

\begin{figure}[t]
\begin{center}
% \hspace*{-0.1\textwidth}
 \includegraphics[width=0.45\textwidth]{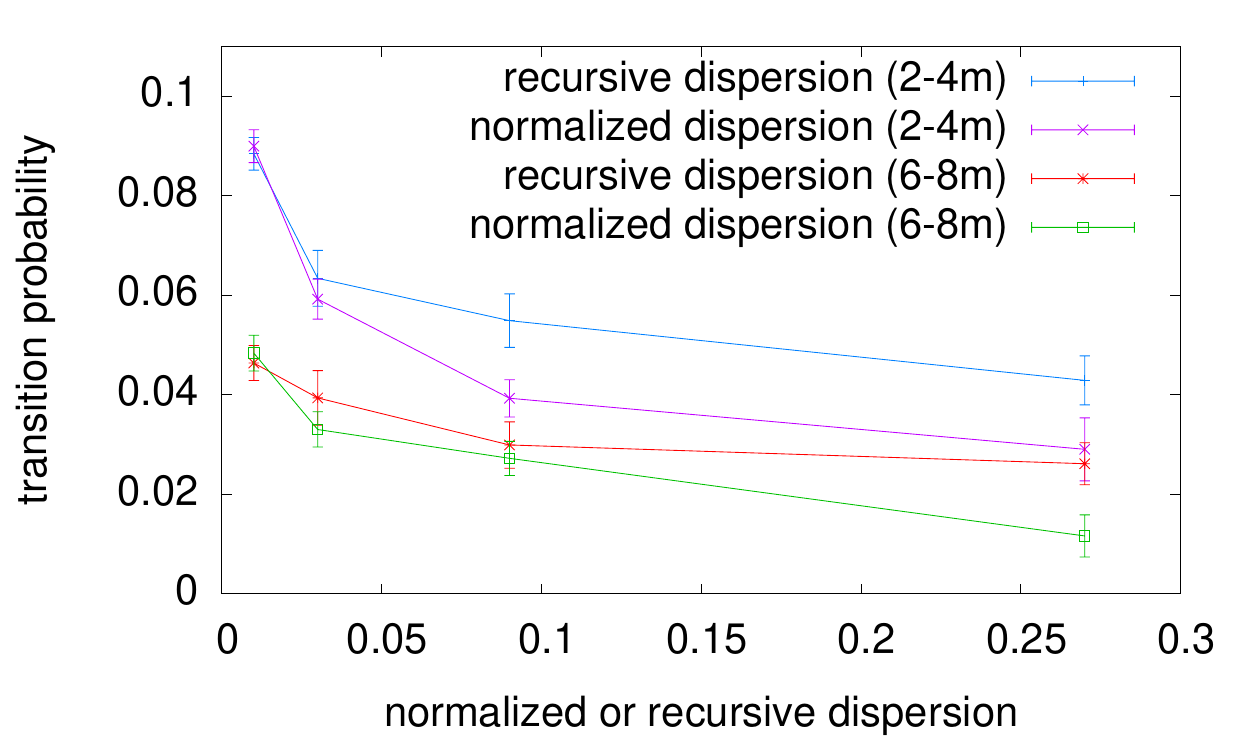}
\caption{
 \label{fig:ja-disp}
% \footnotesize
For different subsets of users, we can evaluate their
{\em transition probability} from the status 
`in a relationship' to the status `single' over a 60-day period.
These transition probabilities are shown as a function of
both normalized and recursive dispersion, separately for relationships
that are 2-4 months old and for relationships that are 6-8 months old.
% (The curves for relationships that are 4-6 months old, not shown,
% interpolate between these two.)
The transition probabilities decrease monotonically, and by
significant factors, for users with high
normalized or recursive dispersion to their respective partners.
\rs \rs 
}
\end{center}
\end{figure}

\begin{figure}[t]
\begin{center}
% \hspace*{-0.1\textwidth}
 \includegraphics[width=0.45\textwidth]{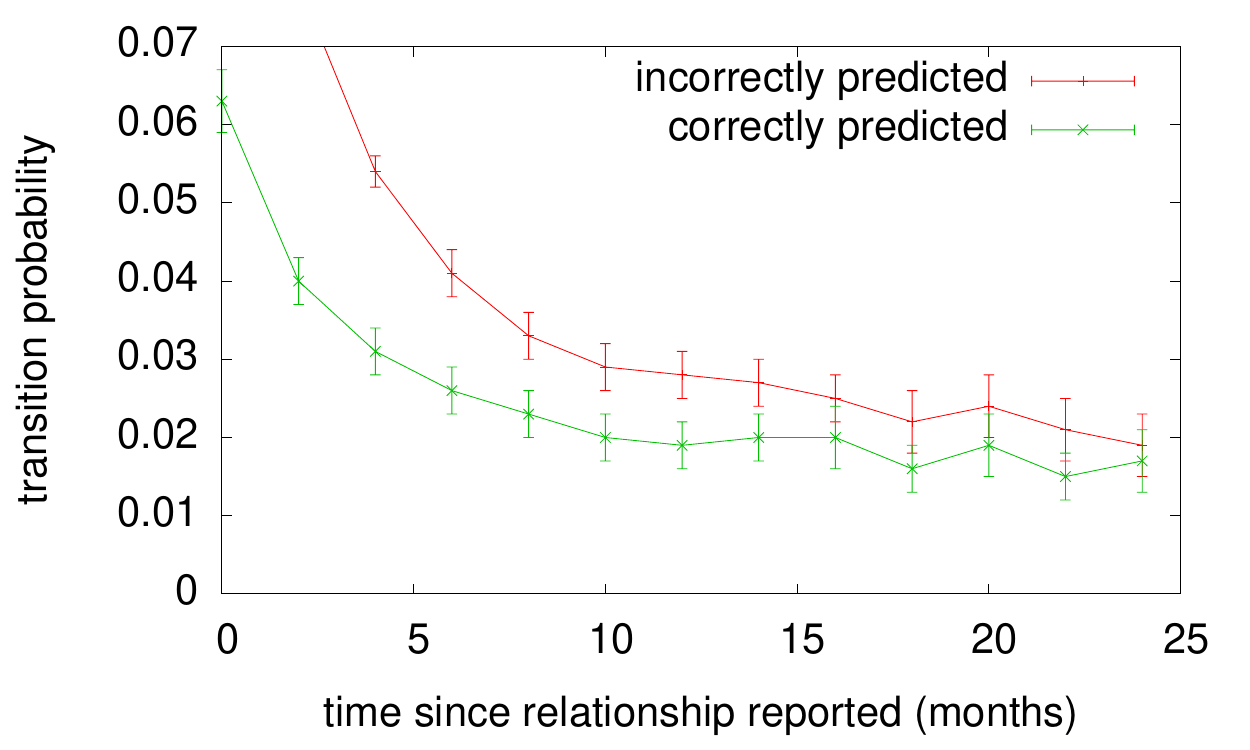}
\caption{
 \label{fig:plot-trans01}
% \footnotesize
We plot the transition probability from `in a relationship' to
`single' as a function of the age of the relationship,
separately for relationships on which recursive dispersion 
correctly identifies the partner and those on which it does not.
The transition probability is significantly higher when 
recursive dispersion fails to recognize the partner.
\rs \rs
}
\end{center}
\end{figure}

Next, we ask how these measures change over time in 
the period leading up to a change in relationship status.
For this purpose, we take a random sample of $29000$ married users
whose relationship went through a succession of stages that were all
reported on Facebook --- first in a relationship, next engagement, 
and then (after a period
of at least a month of engagement) marriage.  For each user $u$ in the sample,
we compute embeddedness as well as normalized
and recursive dispersion for all users in $G_u$ as a function of time,
declaring time $0$ for $u$ to be the moment at which $u$'s marriage was
reported.  For a given network measure $m$,
we then let $f_m(t)$ denote the fraction of instances in 
which $u$'s (future or current) spouse scores highest with respect to $m$.
We see (Fig. \ref{fig:over-time-perf})
that normalized and recursive dispersion rise quickly to the point of
marriage, while embeddedness not only has lower performance but 
also rises more slowly.
When both embeddedness and recursive dispersion 
eventually identify the spouse correctly, recursive dispersion does
so an average of approximately
80 days sooner; in the analogous comparison between 
normalized and recursive dispersion, the latter identifies the spouse
an average of approximately 10 days sooner.

Finally, we consider the persistence of relationships
\cite{le-rel-dissolution}.
Given the effectiveness of these measures in detecting
partners, is it also the case that partnerships that are more strongly
identified by the measures are also more `robust,'
in the sense that they are more likely to persist over time?
We address this question by considering those users in our sample who listed
themselves as being in a relationship (rather than married or
some other relationship status, a set of roughly 400,000 individuals), 
and seeing which of them
lists their relationship status as `single' 60 days later.
We perform these analyses on subsets of the data on which all
relationships have roughly the same age since they were reported
on Facebook, and on which all users were approximately the same age,
to separate these effects from the underlying structural parameters.
We find (Fig. \ref{fig:ja-disp}) that a user whose partner has a
high normalized or recursive dispersion is significantly less
likely to transition to `single' status over this time period.

We can view the persistence of relationships
in a closely related way by comparing relationships on which 
recursive dispersion correctly identifies the partner to 
those on which it does not.
We find (Fig. \ref{fig:plot-trans01}) that relationships on which
recursive dispersion fails to correctly identify the partner 
are significantly more likely to transition to `single' status over
a 60-day period. 
This effect holds across all relationship ages and
is particularly pronounced 
for relationships up to 12 months in age;
here the transition probability is roughly 50\% greater
when recursive dispersion fails to recognize the partner.

%END

%% == end of paper:

%% Optional Materials and Methods Section
%% The Materials and Methods section header will be added automatically.

% \omt{

%% Enter any subheads and the Materials and Methods text below.

\newcommand{\mhdr}[1]{\paragraph*{\bf #1.}}

\section{Beyond Immediate Neighborhoods}
All of the network measures discussed above are based on the immediate
1-hop neighborhoods of individuals. It is interesting to consider how 
accurate more expansive methods might be, if they take
the broader structure of the network into account.  
Because many individuals have 2-hop neighborhoods with hundreds
of thousands of nodes, doing this is computationally
challenging, and we must come up with heuristics to make it 
feasible.

Our approach is to take a single structural measure, recursive 
dispersion, and 
filter down to an individual's top 20 friends as ranked by this metric. 
We then compute the network measures discussed above in the (1-hop)
neighborhoods of each of these 20 people.
To evaluate a given friend $v$ as the potential partner of $u$, we
can then use the measures computed in $u$'s neighborhood and also in $v$'s.
We find that the simple heuristic of taking $u$'s top 20 friends 
with respect to $rec(u,v)$, and then ranking them by 
$\min(rec(u,v),rec(v,u))$,
improves performance by about $6\%$ to $0.534$.
This performs almost as well as more complex models, and confirms the
intuitive result that
relationship partners are best found by looking for pairs of people who have
high scores in both directions.

% \begin{materials}
\section{Conclusion}

Understanding the structural roles of significant people in 
on-line social network neighborhoods is a broad question that
requires a combination of different approaches.
Here we have considered this issue in the context of 
romantic partners, and have identified a novel network measure,
dispersion, that provides a powerful method for recognizing
the structural positions occupied by such partners from network data alone.

Drawing on the theory of social foci \cite{feld-foci}, we have
argued that dispersion is a structural means of capturing the notion 
that a friend spans many contexts in one's social life --- either
because they were present through multiple life stages, or because they 
have been systematically introduced into multiple social circles.
This suggests why it is not only spouses or romantic partners who 
exhibit high dispersion, but also family members --- 
dispersion identifies people who span foci.

There are several applications where
measures based on dispersion may play a role.
For estimating which types of content from friends will be most
engaging to a user \cite{backstrom-wsdm13}, 
identifying individuals with this 
focus-spanning property can be useful for assessing
their properties both as producers and consumers of content
from different parts of a user's network neighborhood.
And for organizing neighborhoods into distinct clusters or circles
\cite{mcauley-social-circles,min-social-roles-cscw13},
dispersion can help identify `hard-to-categorize'
individuals who may need manual annotation from a user ---
high dispersion arises precisely because a friend doesn't naturally
fit into the obvious categories

Beyond these specific applications, our measures suggest
new perspectives on basic questions in social network analysis.
Overall, the notion that our mutual friends with a person may
be clustered in a single context or may alternately span multiple contexts
offers a novel type of trade-off in the study of tie strength.
Certain important types of strong ties ---
including romantic and family relations ---
connect us to people who belong to multiple
parts of our social neighborhood, producing a set of shared friends
that is not simply large but also diverse, spanning disparate
portions of the network and hence correspondingly sparse in
their internal connections.
In this way, the notion of dispersion
combines concepts of network 
closure\cite{granovetter-embeddedness,coleman-social-capital}
(since there must be mutual network neighbors to bridge) and 
brokerage between groups\cite{burt-struct-holes-book}
(since the two ends of a link with large dispersion
are jointly acting as brokers between disconnected
mutual neighbors).  
The success of the measures resulting from this notion
suggests some of the ways in which closure and brokerage
are intertwined in the structure of strong ties.

The analysis shows how these classes of strong ties produce
an extremely clear structural signature, but subtle network
measures different from the standard formulations 
are needed to extract this signature.
Crucial aspects of our everyday lives may be encoded in the network
structure among our friends, provided that we look at this structure
under the right lens.

% \end{materials}

% }

% \omt{
\xhdr{Acknowledgments}
We thank Matt Brashears, Ben Cornwell, Lillian Lee, 
Michael Macy, Cameron Marlow,
Sendhil Mullainathan, Brian Rubineau, and Johan Ugander
for valuable discussions,
and the Simons Foundation and NSF for support.
% This work was supported in part by the Simons Foundation and NSF.
% This research has been
% supported in part by grants from
% the Simons Foundation and from
% the National Science Foundation.
% }

%% Optional Appendix or Appendices
%% \appendix Appendix text...
%% or, for appendix with title, use square brackets:
%% \appendix[Appendix Title]

%% PNAS does not support submission of supporting .tex files such as BibTeX.
%% Instead all references must be included in the article .tex document. 
%% If you currently use BibTeX, your bibliography is formed because the 
%% command \verb+\bibliography{}+ brings the <filename>.bbl file into your
%% .tex document. To conform to PNAS requirements, copy the reference listings
%% from your .bbl file and add them to the article .tex file, using the
%% bibliography environment described above.  

%%  Contact pnas@nas.edu if you need assistance with your
%%  bibliography.

% Sample bibliography item in PNAS format:
%% \bibitem{in-text reference} comma-separated author names up to 5,
%% for more than 5 authors use first author last name et al. (year published)
%% article title  {\it Journal Name} volume #: start page-end page.
%% ie,
% \bibitem{Neuhaus} Neuhaus J-M, Sitcher L, Meins F, Jr, Boller T (1991) 
% A short C-terminal sequence is necessary and sufficient for the
% targeting of chitinases to the plant vacuole. 
% {\it Proc Natl Acad Sci USA} 88:10362-10366.

%% Enter the largest bibliography number in the facing curly brackets
%% following \begin{thebibliography}

%  \bibliographystyle{acm-sigchi}
%  \bibliography{n,extra-refs}

\section{Appendix: Mathematical Properties of the Recursive Dispersion}
\label{sec:rec-disp}

\omt{
$$x_v \longleftarrow 
  \frac{    \sum_{w \in C_{uv}} x_w^2 
        + 2 \sum_{s, t \in C_{uv}} d_v(s,t) x_s x_t}
       {emb(u,v)}.$$
}

This section provides some motivation and further mathematical detail 
related to the recursive dispersion, to help isolate the
ingredients that make up its functional form.

First, fix a distance function $d_v$ on the graph $G - \{u, v\}$.
We first note that if we assign a value $x_s = 1$ to each node $s$
in $G_u$, then $\sum_{s, t \in C_{uv}} d_v(s,t) x_s x_t$ is precisely
the dispersion of $v$, since it is a sum over all pairwise distances in 
$C_{uv}$. Hence, 
$$\frac{\sum_{s, t \in C_{uv}} d_v(s,t) x_s x_t}{emb(u,v)}$$
is the normalized dispersion.

The premise underlying the recursive dispersion is to elevate $x_v$
when $u$ and $v$ act as 
intermediaries between many node pairs $s$ and $t$
that, recursively, have large values of $x_s$ and $x_t$.
A simple way to carry out this idea would be to define an iteration
in which $x_v$ is updated to be 
\begin{equation}
x_v ~ \longleftarrow ~ 
   \frac{\sum_{s, t \in C_{uv}} d_v(s,t) x_s x_t}{emb(u,v)},
\label{eq:rec1}
\end{equation}
directly using the functional form of the normalized dispersion.
The problem with this approach on the graphs $G_u$ that we have
in the data is that the sum in the numerator is equal to $0$ for 
many nodes $v$.  This in turn means that even more nodes will acquire a
$0$ value in subsequent iterations; the end result is
that very few nodes end up with a positive value $x_v$, and
this would hurt the performance in identifying partners.

As a result, it is useful to have a mechanism that continuously
introduces non-zero weight into the system.
As a working guideline for how to do this,
we would like the first iteration to produce values $x_v$
whose sorted order agrees with the order of values according
to normalized dispersion, so that
subsequent iterations can be viewed as building from this measure.
We would also like additional terms in the numerator to 
be quadratic, to match the quadratic degree of the existing terms
in the numerator of (\ref{eq:rec1}).
Adding $\sum_{w \in C_{uv}} x_w^2$ to the numerator achieves
these two properties in a simple way --- it is quadratic, and
in the first iteration it is equal to $emb(u,v)$, and hence
is canceled by the denominator.
This gives us 
the functional form that we use as the recursive dispersion:
\begin{equation}
x_v \longleftarrow 
  \frac{    \sum_{w \in C_{uv}} x_w^2 
        + 2 \sum_{s, t \in C_{uv}} d_v(s,t) x_s x_t}
       {emb(u,v)}.
\label{eq:rec2}
\end{equation}

Depending on the structure of $G_u$, the right-hand side of
(\ref{eq:rec2}) can simplify in several instructive ways.
In what follows, we will also write $emb(u,v)$ in an equivalent
form as $|C_{uv}|$.
\begin{itemize}
\item 
First, suppose $G_u$ has the property that $d_v(s,t) = 1$ whenever
$s$ and $t$ belong to the set $C_{uv}$.
For example, with our distance function $d_v$,
this would hold if the graph $G_u - \{u\}$ 
had no cycle of length
less than $5$; in this case, for any $v$ and any $s, t \in C_{uv}$,
there could be no common neighbor of $s$ and $t$ other than $u$ and $v$,
and so $d_v(s,t) = 1$.
In this case, the iteration in (\ref{eq:rec2}) becomes
\begin{equation}
x_v \longleftarrow
  \frac{    \sum_{w \in C_{uv}} x_w^2
        + 2 \sum_{s, t \in C_{uv}} x_s x_t}
       {emb(u,v)}
= \frac{\left(\sum_{w \in C_{uv}} x_w\right)^2}{|C_{uv}|}
\end{equation}
and so the values evolve according to an iteration that simply sums
values at neighboring nodes and then squares the sum (with
normalization).
\item
Alternately, suppose $G_u$ has the property that $d_v(s,t) = 0$
for all pairs of nodes $s$ and $t$ that belong to the same set $C_{uv}$.
For example, again using our distance function $d_v$, this would be true 
if every pair of nodes at distance $2$ in $G_u$ had at least two
common neighbors other than $u$.
In this case, if in a given iteration we have $x_s = c$ for all $s$,
then in the next iteration
\begin{equation}
x_v \longleftarrow
  \frac{    \sum_{w \in C_{uv}} x_w^2}
       {emb(u,v)}
= \frac{ c^2 |C_{uv}| }{ |C_{uv}| }
= c^2,
\end{equation}
and so all values remain equal to each other.  Thus, for such a $G_u$,
all nodes have the same recursive dispersion.
\end{itemize}

These two special cases represent different conceptual extremes, and
most graphs will lie in between.  
% Given the complicated structure of the iterations, 
It is an interesting open question
to understand the convergence properties of recursive dispersion in
arbitrary graphs.

\end{document}